\documentclass[aps,prl,preprint,showpacs,groupedaddress]{revtex4}

\usepackage[dvips]{graphicx}
\begin{document}
\title{Rocking motion induced charging of $C_{60}$ on $h$-BN/Ni(111)}

\author{M. Muntwiler$^{1}$, W.  Auw\"arter$^{1}$, A.P. Seitsonen$^{2}$, J.~Osterwalder$^{1}$ and T. Greber$^{1}$}

\affiliation{$^{1}$Physik-Institut, Universit\"at Z\"urich, 
Winterthurerstrasse 190, CH-8057 Z\"urich, Switzerland} 
\affiliation{$^{2}$ Institut f\"ur Physikalische Chemie, Universit\"at Z\"urich, 
Winterthurerstrasse 190, CH-8057 Z\"urich, Switzerland}

\begin{abstract} 
One monolayer of $C_{60}$ on one monolayer of hexagonal 
boron nitride on nickel is investigated by photoemission.
Between 150 and 250~K the 
work function decreases and the binding energy of the highest occupied molecular 
orbital (HOMO) increases by $\approx$100~meV. In parallel, the occupancy of
the, in the cold state almost empty, lowest unoccupied molecular orbital (LUMO) changes by $0.4\pm 0.1$ electrons.
This charge redistribution is triggered by onset of molecular rocking 
motion, i.e. by orientation dependent 
tunneling between the LUMO of $C_{60}$ and the substrate.
The magnitude of the charge transfer is large and cannot be explained within 
a single particle picture. It is proposed to involve electron-phonon 
coupling where $C_{60}^{-}$ polaron formation leads to electron self-trapping. 
 
\end{abstract}
\maketitle

In the emerging field of molecular electronics $C_{60}$ 
plays a pivotal role for the exploration and the test of new 
concepts.
The separation of energy scales in inter- and intramolecular bonding
gives rise to structural phase transitions that leave $C_{60}$ intact.
This is e.g. seen with diffraction experiments where orientational 
ordering of $C_{60}$ was inferred from 
changes of the unit cell size
\cite{Heiney:1991,Benning:1993,Goldoni:1996}.
Furthermore, $C_{60}$ compounds have narrow electron bands \cite{Yang:2003}, large on-site Coulomb 
repulsion and intramolecular phonons whose frequencies compare with the 
intermolecular electron hopping rates.
The observation of high $T_{c}$ superconductivity in alkali metal 
doped fullerenes \cite{Hebard:1991} or antiferromagnetism 
\cite{Takenobu:2000} emphasizes strong correlation 
effects.
On this background a rich variety of thermodynamic phases emerges where 
in view of applications metal insulator transitions are particularly 
interesting. 

Here we report on the change of charge state of $C_{60}$ that occurs in parallel 
to an  orientational order-disorder 
transition.
The underlying substrate of the $C_{60}$ monolayer plays a crucial role: It 
provides the charge by tunneling, and it does not bind the 
$C_{60}$ molecules strongly.   
$h$-BN/Ni(111) is an ultimately thin, atomically flat, metal-insulator interface. 
Its
geometry and electronic structure is well known from experiments 
\cite{Nagashima:1995,AuwŠrter:1999} and 
density functional theory \cite{Grad:2003}.
Compared to $C_{60}$/metal 
interfaces \cite{Rudolf:1999,Kiguchi:2003}, the substrate-induced doping, i.e. charge transfer 
on $C_{60}$, is much lower which allows to investigate the insulator limit.
For the description of the $C_{60}$/$h$-BN/Ni(111) junction we borrow concepts 
from non-adiabatic gas surface reactions \cite{Greber2:1997}.
Charge transfer onto a molecule occurs when 
the vertical affinity level is degenerate with the Fermi level of the 
substrate.
Provided that the tunneling electron resides long enough on the 
molecule, the molecule evolves towards its new equilibrium coordinates, 
i.e. approaches the adiabatic affinity level.
The electronic 
system has to pay the energy for this Jahn-Teller distortion and 
can therefore not be described in a one-electron picture.

\begin{figure}[t]
\begin{center}
  \includegraphics[width=6cm]{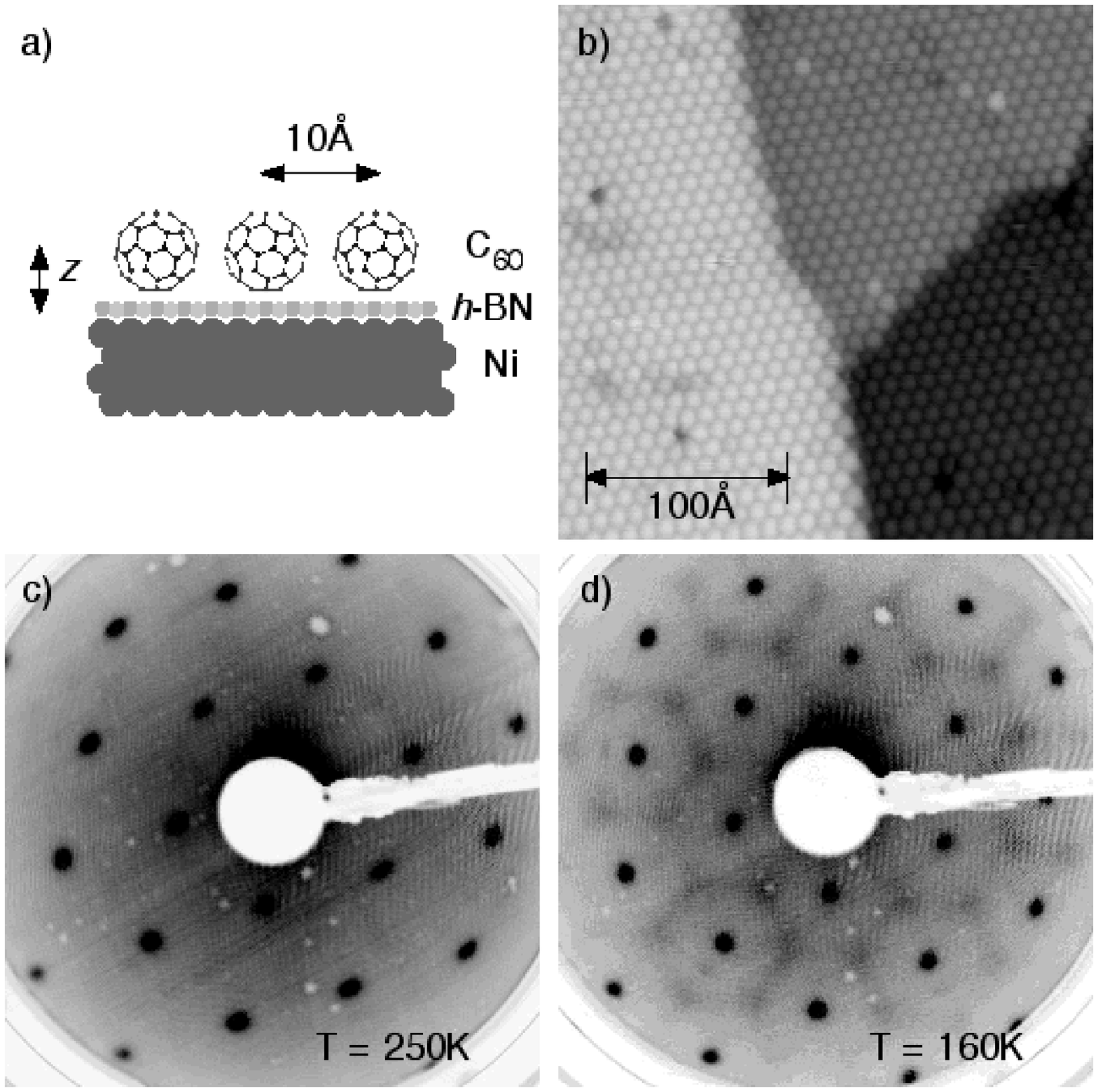}
  \end{center}
  \caption{The investigated system $C_{60}/h$-BN/Ni(111). a) Sketch of 
  the interface. b) STM picture of the well ordered monolayer of $C_{60}$. The three grey levels 
indicate three different terraces on the substrate. c,d) LEED patterns 
($E=23.5$÷eV): (4÷x÷4) at 
250÷K and quasi $(4\sqrt{3}÷{\rm x}÷4\sqrt{3}) {\rm R 30^{\circ}}$ 
at 160÷K.}
  \label{f1}
\end{figure}

The experiments were performed in a modified VG ESCALAB 220 
photoelectron spectrometer 
\cite{Greber:1997} with an attached scanning tunneling microscopy 
(STM) chamber \cite{AuwŠrter:2003}.
The $h$-BN/Ni(111) interfaces were produced using the recipe by Nagashima 
et al. \cite{Nagashima:1995,AuwŠrter:1999}.  
$C_{60}$ monolayers were prepared by annealing multilayers.
On a linear heating 
ramp (0.15 K/s) $C_{60}$ desorbs from the 
multilayer at 475÷K and from the monolayer at 520÷K. 
From this it is inferred that
$C_{60}$ is weakly bound on $h$-BN/Ni(111) ($\approx 1.5$~eV) 
\cite{TDS}.
At room temperature $C_{60}$ wets the substrate and forms a commensurate hexagonally close packed
(4~x~4) structure with a $C_{60}$ nearest neighbor 
distance of 10.0 \AA~(see Figure \ref{f1}).
The STM picture in Figure \ref{f1}b) shows how perfect the $C_{60}$ 
/$h$-BN/Ni(111) junction grows. 

Figures \ref{f1} c) and d) are low energy electron diffraction (LEED) 
data ($E=23.5$÷eV) at 250 and 160~K, respectively.   
From comparison of the two it is seen that the unit cell size 
increases from a (4x4) with one $C_{60}$ molecule to a quasi 
($4\sqrt{3}÷{\rm x}÷4\sqrt{3} {\rm R 30^{\circ}}$) unit cell with 3 
$C_{60}$ molecules when the sample is cooled to
160÷K. 
This is in line with the findings of Benning et al. 
\cite{Benning:1993} and Goldoni et al. \cite{Goldoni:1996}, who found for $C_{60}$ 
multilayers in the same temperature range a (1x1) to 
(2x2) surface lattice transformation.
This phase transition corresponds to an
order disorder transition which occurs in bulk $C_{60}$ at 
$\approx$÷250÷K \cite{Heiney:1991}.

\begin{figure}[h]
\begin{center}
  \includegraphics[width=8cm]{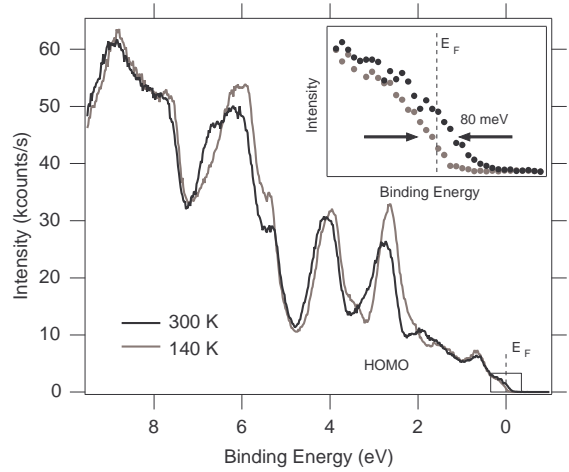}
  \end{center}
  \caption{He I${\alpha}$ photoemission spectra of one monolayer of 
  $C_{60}$ on $h$-BN/Ni(111) for room temperature and 160 K. 
  The inset shows the strong shift of the photoemission leading edge that is due to the 
population of the LUMO at high temperatures.}
  \label{f2}
\end{figure}

In Figure \ref{f2} He I$\alpha$ normal emission photoemission spectra for 
300 and 160~K are shown. 
Due to the small mean free path of the photoelectrons the molecular orbitals of 
$C_{60}$ dominate the spectra and the underlying 
substrate contributes only weakly. 
The bump at 0.6~eV binding energy reflects the Ni $d$-band.
The $C_{60}$ derived features are strongly temperature dependent. 
The highest
occupied molecular orbital (HOMO) shifts from 2.75 up to 2.64~eV in 
going from 250~K to 160~K.
Furthermore, the orbital peaks sharpen and the HOMO increases in intensity.
From the large HOMO binding energy and the fact that there is no strong 
LUMO contribution which would 
produce a peak at the Fermi level we see that this system is an 
example for poor screening, i.e. high on-site Coulomb repulsion or
large HOMO-LUMO separation in photoemission \cite{Rudolf:1999}.
The work function increases parallel to the HOMO-shift from 4.25 to 
4.37~eV.
This indicates a charge redistribution in the interface and that the 
molecular orbitals are mainly bound to the vacuum level. 

The inset of Figure~\ref{f2} shows the spectra in the vicinity of the 
Fermi level. 
Intriguingly, a  80~meV down shift of the leading edge of the photoemission 
is observed in going from room temperature to 160 K.
Such a shift is beyond instrumental uncertainties since the  
energy resolution was better than 60~meV.
A careful comparison with the Fermi edge as measured on a Ag 
polycrystal indicates that the room temperature leading 
edge lies above $E_{F}$.
As we show below it reflects a shift of the high binding energy wing of the 
LUMO parallel to the HOMO.
The difference between the room temperature and the 160~K 
spectrum in the inset of Figure \ref{f2} corresponds to 4.5 \% 
of the HOMO intensity at 300 K.
This intensity can be assigned to the charge difference on the 
LUMO at high and low temperature, respectively.

\begin{figure}[h]
\begin{center}
  \includegraphics[width=8cm]{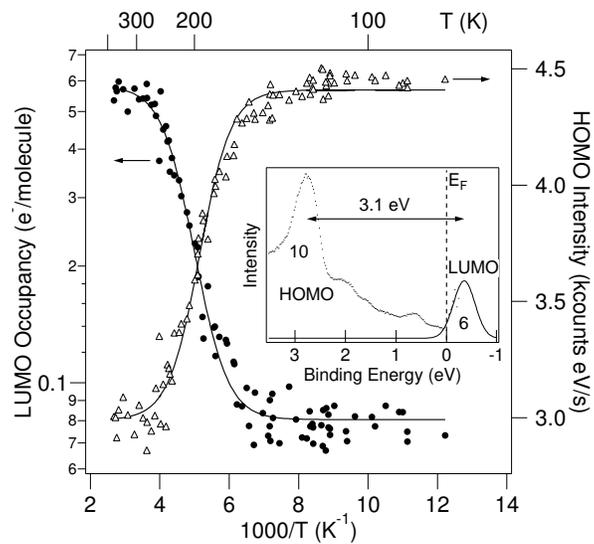}
  \end{center}
  \caption{Occupancy of the LUMO (solid 
  circles) and normal emission HOMO intensity (open triangles) as a 
  function of reciprocal temperature. The solid lines are guides to the eye.
  The inset shows the extrapolated LUMO from the 300~K data in Figure~\ref{f2}.}
  \label{f3}
\end{figure}

To get a more quantitative measure of the LUMO occupancy a one 
parameter fit was applied to the data.
The fit procedure asserts the same Gaussian spectral shape as that of the HOMO and a
degeneracy-derived weight of 10 for the $h_u$ HOMO and 6 for a
fully occupied
$t_{1u}$ LUMO \cite{Cepek:2000}. 
The Gaussian for the LUMO is fitted to the spectrum that was
divided by the
corresponding experimental Fermi function (see inset of 
Figure \ref{f3}).
The only fit parameter is the HOMO-LUMO energy difference 
\cite{HOMO-LUMO},
and the fit is performed in a region from the Fermi energy up to 200 meV 
above $E_F$.
In order to obtain the LUMO occupancy the Gaussian for the LUMO is 
multiplied with the Fermi function and integrated.
Figure \ref{f3} shows this LUMO 
occupancy versus the reciprocal 
temperature. 
It increases by a factor of $7\pm 1$ in going from 150 to 
300~K \cite{fit}.
Though these data are recorded during cool down of the sample, the 
effect is reversible with a hysteresis $<$ 50~K.
From the fit the LUMO occupancy changes between room temperature and 
160 K by $\approx$ half an electron. 
This is in good agreement with 0.45 electrons \cite{reproducibility} as 
indepenently derived from the difference of 
the spectra near the Fermi energy (inset of Figure \ref{f2}) and the comparison with the HOMO 
intensity.
In Figure \ref{f3} we show as well the intensity of the HOMO.
It decreases in a non Debye-Waller fashion with increasing temperature 
and indicates changes in molecular orientation as reflected in ultraviolet photoelectron diffraction (UPD) 
\cite{Osterwalder:1996}. 
The transition temperature coincides with that of the change in charge state.
This is a clear indication that the charge state is closely 
related to the molecular orientation. Since the spectra were 
recorded in normal emission it can be inferred that the 
change in orientation must contain rocking motion, i.e. 
rotational axes that do not coincide with the surface normal. 

How can we explain the change in charge transfer with temperature?
In Bardeen's tunneling picture the charge transfer 
between the substrate and the adsorbate is related to the overlap of 
the wave functions at the Fermi level \cite{Bardeen:1961}.
The charge on the $C_{60}$ molecules thus reflects a finite tunneling 
probability of conduction electrons across the 
$h$-BN layer onto the LUMO of $C_{60}$.
The residual occupancy of 0.08(2) electrons at low temperatures indicates that 
electron tunneling from the substrate to the molecule is still 
enabled, though at a low rate.
At temperatures above 250~K the charge transfer begins to saturate.
From this non Arrhenius behavior a model where the charge 
transfer is induced by thermally excited electrons alone, has to be refuted. 
However, the increase of the LUMO occupancy with temperature is
related to the onset of molecular rocking motion in the $C_{60}$ monolayer
(see Figure \ref{f3}) and therefore a molecular orientation dependent wave function overlap 
with the substrate gives a natural explanation for the effect.
The onset of molecular rocking motion is in line with the known disorder transition in $C_{60}$ multilayers 
\cite{Benning:1993,Goldoni:1996}, the LEED patterns in Figure \ref{f1}, X-ray 
photoelectron diffraction (XPD) and Ultraviolet Photoelectron 
Diffraction (UPD) (also see Figure \ref{f3}) patterns.
Furthermore, the XPD patterns  (not shown) indicate that the $C_{60}$ molecules do not bind via 
the pentagons but rather via hexagons to the substrate, as it is e.g. 
the case for $C_{60}$ on Cu(111) \cite{Fasel:1996}.

In order to justify the molecular rocking driven model, the electronic structure of
$C_{60}$ has to deviate from spherical symmetry.
For this purpose density functional calculations were performed within the Kohn-Sham
scheme. As the exchange correlation functional we employed the
Generalized Gradient Approximation (GGA) of Perdew, Burke and
Ernzerhof \cite{DFT}.
The
Kohn-Sham orbitals were expanded either in a Gaussian basis set TZVP
in the TURBOMOLE code \cite{Turbomol} or in plane waves with a cutoff energy of 40
Ry. In the latter case the action of the core electrons on the valence
electrons was replaced with norm-conserving pseudo potentials, and the
length of the supercell was 21.2 {\AA}.

\begin{figure}[h]
\begin{center}
  \includegraphics[width=8.5cm]{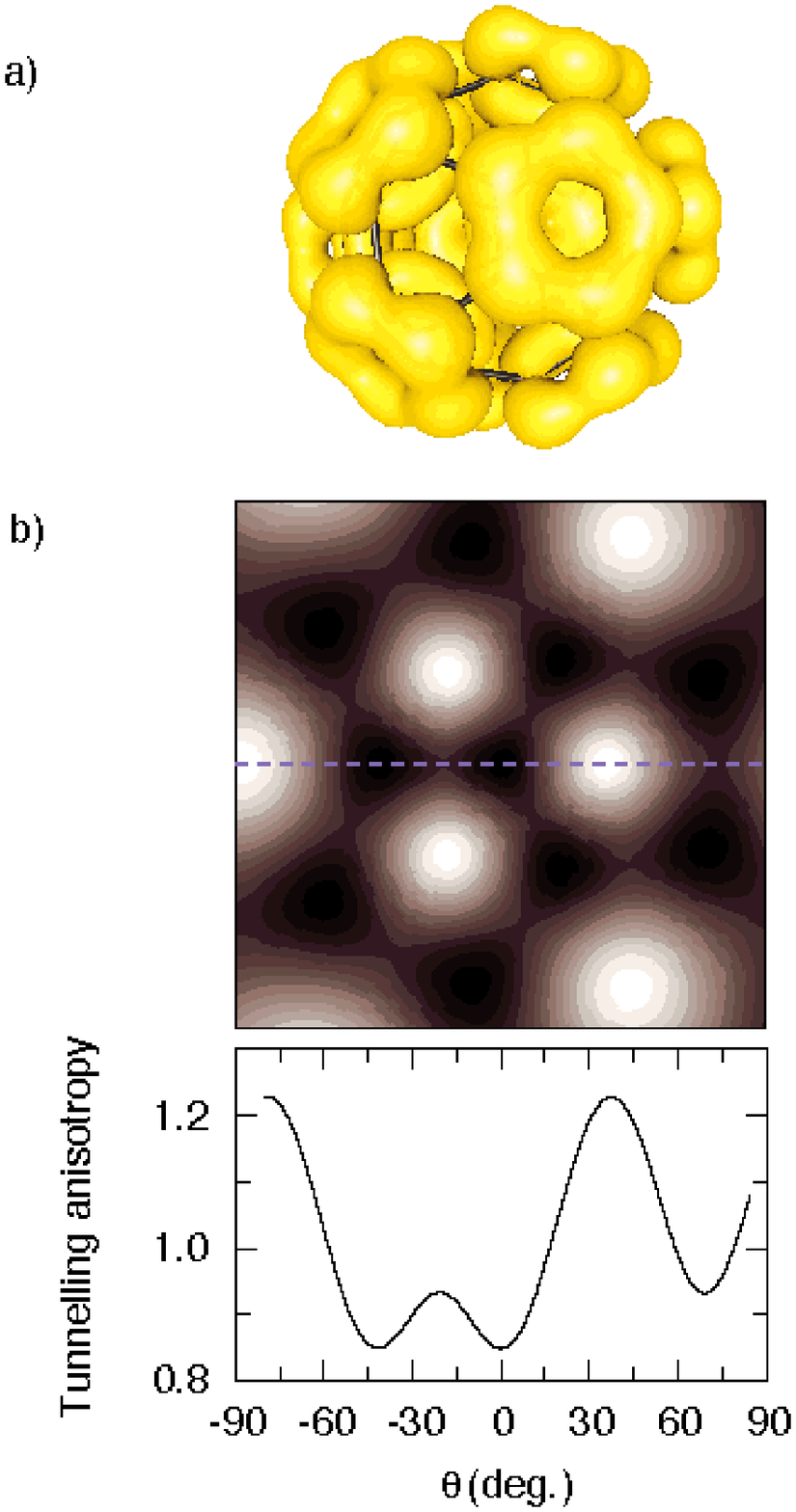}
  \end{center}
  \caption{Deviation of the LUMO of $C_{60}$ from spherical 
  symmetry.
  a) Constant electron density isosurface.
  b) Stereographic projection of the integral electron density outside 
  planes placed
  5~\AA~ away from the center of $C_{60}$. The low density parallel to hexagons 
  (center of the plot) and the high density 
  parallel to pentagons is apparent. The bottom 
  panel shows the anisotropy of this electron density that is 
  related to the tunneling rate between the substrate and the 
  molecule along the dashed $\theta$ line.
  The tunneling anisotropy of 1 corresponds to the average of all 
  molecular orientations.}
  \label{f4}
\end{figure}

Figure \ref{f4} shows results of density functional calculations of 
$C_{60}$. 
In Figure \ref{f4}a) the degenerate LUMO is shown as a constant electron density 
isosurface. 
Clearly, the LUMO electron density is largest on the pentagons, 
while the hexagons have a low density.
Together with the fact that at low temperatures $C_{60}$ does not 
expose 
pentagons towards the substrate this gives a consistent picture: The 
overlap between the LUMO and the substrate, i.e. the tunneling rate 
increases if the molecules rock away from the equilibrium position 
and correspondingly more electrons tunnel into the LUMO. 
This overlap is shown in Figure \ref{f4}b) as a function of the molecular 
orientation relative to the surface normal.
The grey scale plot gives the integral of the LUMO in the 
half space $z > z_{o}$ with $z_{o} = 5÷$\AA~ away from the center of $C_{60}$ as a function of 
the molecular orientation relative to this plane.
The average probability to find an electron outside a plane $z > z_{o}$ i.e. 
the mean of the integrals over all orientations $\int |\Psi_{LUMO}(z>z_{o})|^{2}$ is $ 0.007÷\int |\Psi_{LUMO}|^{2}$  
and the anisotropy that we call 
tunneling anisotropy may be seen in the bottom panel of Figure 
\ref{f4}b).
The tunneling anisotropy slightly depends on $z_{o}$, though for reasonable 
$z_{o} \ge 5 $ \AA~ it never reaches values that could explain 
quantitatively the
strong temperature dependence of the LUMO occupancy.
From the experiment the LUMO occupancy ratio between room temperature and low 
temperature is $\approx 7$, while a tunneling anisotropy of 0.75 could 
explain a ratio of 1.3 only.
In other terms this theory explains an occupancy change of 0.01 
electrons only, which is inconsistent with observation.   

Thus we have to refine the orientation dependent tunneling 
picture.
It is unlikely that the substrate wave function and/or distance 
induced changes in the tunneling matrix element account for the strong
{\it amplification} of the effect. 
Also, electron hopping between the $C_{60}$ molecules is not expected 
to change the net charge balance in such a dramatic way.
It is rather an indication 
that the one-electron picture as it is drawn 
from the tunneling anisotropy in Figure \ref{f4}b) breaks down.
The low tunneling rate provides 
substantial time for an electron to stay and to act on the $C_{60}$ 
layer before it 
returns to the substrate. 
For large residence times, comparable with the molecular vibration 
periods, the $C_{60}^{-}$ species start 
to deform and the electron couples strongly to phonons.
In other terms, $C_{60}^{-}$ distorts from the vertical affinity 
$E_{A}^{V}$ towards the adiabatic 
affinity $E_{A}^{A}$.
Our calculations on gas-phase $C_{60}^{-}$ result in a Jahn-Teller distortion energy, i.e.
energy difference between unrelaxed and relaxed $C_{60}^{-}$ 
$E_{A}^{V}-E_{A}^{A}$ of 51 meV.
In the condensed phase we have to consider intramolecular {\it and} extramolecular 
vibrations. Though, their interplay and relative importance is not known 
at the moment.
Intramolecular electron phonon coupling is e.g. observed in photoemission from $C_{60}^{-}$ 
\cite{Gunnarsson:1991}, or recently in inelastic tunneling 
experiments on $C_{60}$/Ag(110), where an electron energy loss of $54$÷
meV was assigned to the excitation of a $H_{g}$ phonon \cite{Pascual:2002}.
Near metallic surfaces the coupling to extramolecular vibrations occurs via the 
image force on the $C_{60}^{-}$ ion, which excites frustrated translational 
vibrations. 
In this image field the occupied LUMO dives below the Fermi level and 
gets trapped since back-tunneling to the substrate is prevented due to the 
lack of empty states.
Together with the onset of Coulomb repulsion this may establish a new 
equilibrium LUMO occupancy.
If indeed such a kind of `polaron' driven self trapping mechanism is 
responsible for the strong amplification of the tunneling 
anisotropy this is a particular example for strong 
electron phonon coupling in such an interface system.
 
In summary we report the observation of a temperature driven change 
in $C_{60}$ molecular orbital binding energies parallel to the work function and 
a strong change in LUMO occupancy of $C_{60}$ on $h$-BN/Ni(111) that coincide 
with the onset of molecular rocking. 
The charge transfer is triggered by an anisotropic 
tunneling matrix element of the  $C_{60}$ LUMO with the underlying 
substrate.
The magnitude of the charge transfer is an indication for 
strong electron-phonon interactions.  
This effect might open the door for the 
application of the orientation dependence of the tunneling across 
molecules for electronic and spintronic switches.

 Fruitful discussions with F. Baumberger and R. Monnier and 
 funding from the Schweizerischen Nationalfonds is 
 gratefully acknowledged.  

\newpage

\begin{thebibliography}{30}
\bibitem{HOMO-LUMO}
The HOMO-LUMO energy difference remains constant ($\approx 3.1$~eV) and is 
close to the value of undoped $C_{60}$.
\bibitem{fit}
The HOMO:LUMO intensity ratio may depend on the molecular orientation. 
If it changes by a 
unrealistically large factor of 2,
we would expect the LUMO occupancy to change by a factor of 3 instead of 7.
\end{thebibliography}
\end{document}